# SOME IMPROVED ESTIMATORS IN SYSTEMATIC SAMPLING UNDER NON-RESPONSE

**Hemant K. Verma, R. D. Singh and Rajesh Singh**

**Department of Statistics, Banaras Hindu University, Varanasi-221005, India**

**Abstract**

In this paper we have considered the problem of estimating the population mean in systematic sampling using information on an auxiliary variable in presence of non – response. Some modified ratio, product and difference type estimators in systematic sampling have been suggested and their properties are studied. The expressions of mean squared error's (MSE's) up to the first order of approximation are derived. An empirical study is carried out to judge the best estimator out of the suggested estimators.

## 1. Introduction

In survey sampling use of auxiliary information can increase the precision of an estimator when study variable y is highly correlated with the auxiliary variable x. Many authors suggested estimators using some known population parameters of an auxiliary variable. [1-5] suggested estimators in simple random sampling.

But in several practical situations, instead of existence of auxiliary variable there exists some auxiliary attributes which are highly correlated with study variable y. In such situations, taking the advantage of point bi-serial correlation between the study variable and the auxiliary attribute, the estimators of parameters of interest can be constructed by using prior knowledge of the parameter of auxiliary attributes. [3] and [6-10] have considered the problem of estimating population mean using point bi-serial correlation between study variable and auxiliary attribute.

The importance of systematic sampling cannot be overemphasized, being one of the sampling schemes most widely used in practice due to its appealing simplicity. The method of systematic sampling first studied by [11] and is widely used in survey of finite populations. Use of auxiliary information in construction of estimators is considered by [12-15].

Systematic sampling is a method of selecting sample members from a larger population according to a random starting point and a fixed, periodic interval. Typically, every "nth" member is selected from the total population for inclusion in the sample population. Systematic sampling is still thought of as being random, as long as the periodic interval is determined beforehand and the starting point is random.

The usual ratio, product and regression estimators of the population mean $\bar{Y}$ based on a systematic sample of size n, under the assumption that the population mean $\bar{X}$ is known, can be respectively defined as

$$\bar{y}_R^* = \frac{\bar{y}^*}{\bar{x}^*}\bar{X} \tag{1.1}$$

$$\bar{y}_P^* = \frac{\bar{y}^*\bar{x}^*}{\bar{X}} \tag{1.2}$$

$$\bar{y}_{lr}^* = \bar{y}^* + b\left(\bar{X} - \bar{x}^*\right) \tag{1.3}$$

Where $b = \frac{s_{xy}}{s_x^2}$, and $\bar{y}^*$, $\bar{x}^*$ are estimators of population mean $\bar{Y}$ (study variables) and $\bar{X}$ (auxiliary variable), respectively, based on the systematic sample of size n unit.

The MSE's of the estimators $\bar{y}_R^*$, $\bar{y}_P^*$ and $\bar{y}_{lr}^*$ are respectively, given by

$$MSE\left(\bar{y}_R^*\right) = \theta\bar{Y}^2\{1 + (n-1)\rho_x\}\left[\rho^{*2} C_Y^2 + (1 - 2K_1\rho^*)C_X^2\right] \tag{1.4}$$

$$MSE\left(\bar{y}_P^*\right) = \theta\bar{Y}^2\{1+(n-1)\rho_X\}\left[\rho^{*2}C_Y^2 + (1+2K_1\rho^*)C_X^2\right] \tag{1.5}$$

$$MSE\left(\bar{y}_{lr}^*\right) = \theta\bar{Y}^2\{1+(n-1)\rho_X\}\left[C_Y^2 - K_1^2C_X^2\right]\rho^{*2} \tag{1.6}$$

where, $\theta = \frac{N-1}{nN}$, $\rho_x = \frac{E(x_{ij}-\bar{X})(x_{ij}-\bar{X})}{E(x_{ij}-\bar{X})^2}$, $\rho_y = \frac{E(y_{ij}-\bar{Y})(y_{ij}-\bar{Y})}{E(y_{ij}-\bar{Y})^2}$

$$\rho = \frac{E(x_{ij}-\bar{X})(y_{ij}-\bar{Y})}{\left(E(x_{ij}-\bar{X})^2(y_{ij}-\bar{Y})^2\right)^{\frac{1}{2}}}, \quad \rho^* = \frac{\{1+(n-1)\rho_Y\}^{1/2}}{\{1+(n-1)\rho_X\}^{1/2}}, \quad K_1 = \rho\frac{C_Y}{C_X}.$$

And $C_Y$ , $C_X$ are the coefficients of variations of study and auxiliary variables respectively.

In this paper we have proposed a general class of ratio, product and difference type estimators for estimating the population mean in systematic sampling using auxiliary information in the presence of non-response. A comparative study is also carried out to compare the optimum estimators with respect to usual mean estimator with the help of numerical data.

## 2. Non Response

Non-response means failure to obtain a measurement on one or more study variables for one or more elements selected for the survey. Let us suppose that a population consists of N units numbered from 1 to n in some order and a sample of size n is to be drawn such that N = nk (k is an integer). Thus there will be k samples each of n units and we select one sample from the set of k samples. Let Y and X be the study and auxiliary variable with respective means $\bar{Y}$ and $\bar{X}$. Let us consider $y_{ij}$ ($x_{ij}$) be the $j^{th}$ observation in the $i^{th}$ systematic sample under study (auxiliary) variable (i=1…k : j=1…n).

We assume that the non-response is observed only on study variable and auxiliary variable is free from non-response. Using [16] technique of sub-sampling of non-respondents, the estimator of population mean $\bar{Y}$, can be defined as

$$\bar{y}^{**} = \frac{n_1 \bar{y}_{n1} + n_2 \bar{y}_{h_2}}{n} \tag{2.1}$$

Where $\bar{y}_{n1}$ and $\bar{y}_{h_2}$ are, respectively the means based on $n_1$ respondent units from the systematic sample of n units and sub-sample of $h_2$ units selected from $n_2$ non-respondent units in the systematic sample. The estimator of population mean $\bar{X}$ of auxiliary variable based on the systematic sample of size n units, is given by

$$\bar{x}^{*} = \frac{1}{n}\sum_{j=1}^{n} x_{ij} \qquad (i = 1...k) \tag{2.2}$$

Obviously, $\bar{y}^{**}$ and $\bar{x}^{*}$ are unbiased estimators. The variance expression for the estimators $\bar{y}^{**}$ and $\bar{x}^{*}$ are, respectively, given by

$$V\left(\bar{y}^{**}\right) = \theta(1 + (n-1)\rho_Y)S_Y^2 + \frac{L-1}{n} K S_{Y2}^2 \tag{2.3}$$

$$V\left(\bar{x}^{*}\right) = \theta(1 + (n-1)\rho_x)S_x^2 \tag{2.4}$$

Where $\rho_Y$ and $\rho_x$ are the correlation coefficients between a pair of units within the systematic sample for the study and auxiliary variables respectively. $S_Y^2$ and $S_x^2$ are respectively, the mean square of the entire group for study and auxiliary variable. $S_{Y2}^2$ be the mean square of non-response group under study variable, K is the non-response rate in the population and $L = \frac{n_2}{h_2}$.

The ratio, product and regression estimators defined in equation (1.1), (1.2) and (1.3) under non-response can be respectively, written as

$$\bar{y}_R^{**} = \frac{\bar{y}^{**}}{\bar{x}^{*}} \bar{X} \tag{2.5}$$

$$\bar{y}_P^{**} = \frac{\bar{y}^{**}\bar{x}^{*}}{\bar{X}} \tag{2.6}$$

$$\bar{y}_{lr}^{**} = \bar{y}^{**} + b\left(\bar{X} - \bar{x}^{*}\right) \tag{2.7}$$

The MSE expression for these estimators are respectively given by

$$MSE\left(\bar{y}_R^{**}\right) = \theta\bar{Y}^2\{1 + (n-1)\rho_x\}\left[\rho^{*2} C_Y^2 + (1 - 2K_1\rho^{*})C_X^2\right] + \frac{L-1}{n}W_2S_{Y2}^2 \tag{2.8}$$

$$MSE\left(\bar{y}_P^{**}\right) = \theta\bar{Y}^2\{1 + (n-1)\rho_x\}\left[\rho^{*2} C_Y^2 + (1 + 2K_1\rho^{*})C_X^2\right] + \frac{L-1}{n}W_2S_{Y2}^2 \tag{2.9}$$

$$MSE\left(\bar{y}_{lr}^{**}\right) = \theta\bar{Y}^2\{1 + (n-1)\rho_x\}\left[C_Y^2 - K_1^2C_X^2\right]\rho^{*2} + \frac{L-1}{n}W_2S_{Y2}^2 \tag{2.10}$$

**3. Proposed improved estimators**

In this section we propose some improved estimators. First, we propose an estimator $t_1$ as

$$t_1 = \bar{y}^{*}\left[\frac{\bar{X} - \alpha\left(\bar{X} - \bar{x}^{*}\right)}{\bar{x}^{*} + \alpha\left(\bar{X} - \bar{x}^{*}\right)}\right] \tag{3.1}$$

Where α is a constant.

$$t_2 = \bar{y}^{*}\left[\frac{\bar{x}^{*} + a\left(\bar{X} - \bar{x}^{*}\right)}{\bar{x}^{*} + b\left(\bar{X} - \bar{x}^{*}\right)}\right]^p \tag{3.2}$$

Where a, b and p are constants.

Adapting [17] estimator in systematic sampling we propose an estimator $t_3$ as:

$$t_3 = \bar{y}^* \left[ 2 - \left( \frac{\bar{x}^*}{\bar{X}} \right)^w \right] \quad (3.3)$$

Where w is a constant.

We propose a difference type estimator $t_4$ as

$$t_4 = K_{41} \bar{y}^* \left[ \frac{\bar{X} - \alpha\left(\bar{X} - \bar{x}^*\right)}{\bar{x}^* + \alpha\left(\bar{X} - \bar{x}^*\right)} \right] + K_{42}\left(\bar{X} - \bar{x}^*\right) \quad (3.4)$$

Where $K_{41}, K_{42}$ and α are constant.

We propose two another improved estimators $t_5$ and $t_6$ as

$$t_5 = K_{51} \bar{y}^* \left[ \frac{\bar{x}^* + a\left(\bar{X} - \bar{x}^*\right)}{\bar{x}^* + b\left(\bar{X} - \bar{x}^*\right)} \right]^p + K_{52}\left(\bar{X} - \bar{x}^*\right) \quad (3.5)$$

Where $K_{51}, K_{52}, a, b$ and p are constant.

$$t_6 = K_{61} \bar{y}^* \left[ 2 - \left( \frac{\bar{x}^*}{\bar{X}} \right)^w \right] + K_{62}\left(\bar{X} - \bar{x}^*\right) \quad (3.6)$$

Where $K_{61}, K_{62}$ and w are constants.

Using the usual procedure we get the expressions for the biases of the above estimators as

$$\text{Bias}(t_1) = \theta\overline{Y}\left[\left(1-3\alpha+2\alpha^2\right)C_1^2 - (1-2\alpha)\rho C_0 C_1\right] \tag{3.7}$$

$$\text{Bias}(t_2) = A\theta C_1^2 - D\theta C_0 C_1 \tag{3.8}$$

$$\text{Bias}(t3) = \overline{Y}\left[-\frac{1}{2}w(w-1)\theta C_1^2 - w\rho C_0 C_1\right] \tag{3.9}$$

$$\text{Bias}(t_4) = \overline{Y}(K_{41}-1) + K_{41}\overline{Y}\theta\left[\left(1-3\alpha+2\alpha^2\right)C_1^2 - (1-2\alpha)\rho C_0 C_1\right] \tag{3.10}$$

$$\text{Bias}(t_5) = K_{51}\overline{Y}\left[1+\theta\left(AC_1^2 - D\rho C_0 C_1\right)\right] - \overline{Y} \tag{3.11}$$

$$\text{Bias}(t_6) = (K_{61}-1)\overline{Y} - K_{61}\overline{Y}\theta\left[\frac{1}{2}w(w-1)C_1^2 + w\rho C_0 C_1\right] \tag{3.12}$$

where $A = \frac{p(p+1)}{2}(1-b)^2 - p^2(1-a)(1-b) + \frac{p(p-1)}{2}(1-a)^2$
and $D = (a - b)p.$

Similarly, the expressions of MSE's of the above estimators are given by

$$\text{MSE}(t_1) = \theta\overline{Y}^2\left[C_0^2 + (1-2\alpha)^2 C_1^2 - 2(1-2\alpha)\rho C_0 C_1\right] + \frac{(L-1)}{n}KS_{Y2}^2 \tag{3.13}$$

Differentiating expression (3.13) with respect to w, we get the optimum value of α ( α*) as-

$$\alpha^* = \frac{1}{2}\left(\rho\frac{C_0}{C_1} - 1\right)$$

$$\text{MSE}(t_2) = \overline{Y}^2\theta\left[D^2C_1^2 + C_0^2 - 2\rho D C_0 C_1\right] + \frac{(L-1)}{n}KS_{Y2}^2 \tag{3.14}$$

Differentiating expression (3.14) with respect to w, we get the optimum value of D ( D*) as-

$$D^* = \rho \frac{C_0}{C_1}.$$

$$MSE(t_3) = \overline{Y}^2 \theta \left[ w^2 C_1^2 + C_0^2 - 2w\rho C_0 C_1 \right] + \frac{(L-1)}{n} K S_{Y2}^2 \tag{3.15}$$

Differentiating expression (3.15) with respect to w, we get the optimum value of w ( w*) as-

$$w^* = \rho \frac{C_0}{C_1}$$

$$MSE(t_4) = K_{41}^2 \overline{Y}^2 A_{41} - 2K_{41} K_{42} \overline{X}\overline{Y} A_{42} - 2K_{41} \overline{Y}^2 A_{43} + K_{42}^2 \overline{X}^2 A_{44} + \overline{Y}^2 + \frac{(L-1)}{n} K_{41}^2 K S_{Y2}^2 \tag{3.16}$$

Where $A_{41} = 1 + \theta\left[\left(3 - 10\alpha + 8\alpha^2\right)C_1^2 + C_0^2 - 4(1-2\alpha)\rho C_0 C_1\right]$

$$A_{42} = \theta\left[-(1-2\alpha)C_1^2 + \rho C_0 C_1\right]$$

$$A_{43} = 1 + \theta\left[\left(1 - 3\alpha + 2\alpha^2\right)C_1^2 - (1-2\alpha)\rho C_0 C_1\right]$$

$$A_{44} = \theta C_1^2$$

And the optimum value of $K_{41}$ and $K_{42}$ are given by

$$K_{41}^* = \frac{A_{43} A_{44}}{A_{41} A_{44} + A_{42}^2 + \frac{(L-1)}{n} K A_{44} \frac{S_{Y2}^2}{\overline{Y}^2}}$$

$$K_{42}^* = \frac{\overline{Y} A_{42} A_{43}}{\overline{X}\left[A_{41} A_{44} + A_{42}^2 + \frac{(L-1)}{n} K A_{44} \frac{S_{Y2}^2}{\overline{Y}^2}\right]}$$

$$MSE(t_5) = K_{51}^2 \overline{Y}^2 A_{51} + K_{52}^2 \overline{X}^2 A_{52} - 2K_{51} K_{52} \overline{X}\overline{Y} A_{53} - 2K_{51} \overline{Y}^2 A_{54} + \overline{Y}^2 + \frac{(L-1)}{n} K K_{51}^2 S_{Y2}^2$$

(3.17)

Where $A_{51} = 1 + \theta\left[C_0^2 + \left(D^2 - 2A\right)C_1^2 - 4D\rho C_0 C_1\right]$

$$A_{52} = \theta C_1^2$$

$$A_{53} = \theta\left[-DC_1^2 + \rho C_0 C_1\right]$$

$$A_{54} = 1 + \theta\left[AC_1^2 - D\rho C_0 C_1\right]$$

And the optimum value of $K_{51}$ and $K_{52}$ is given as

$$K_{51} = \frac{A_{52}A_{54}}{A_{51}A_{52} + A_{53} + \frac{(L-1)}{n}KA_{52}\frac{S_{Y2}^2}{\bar{Y}^2}}$$

$$K_{52} = \frac{\bar{Y}A_{52}A_{54}}{X\left[A_{51}A_{52} + A_{53} + \frac{(L-1)}{n}KA_{52}\frac{S_{Y2}^2}{\bar{Y}^2}\right]}$$

$$MSE(t_6) = K_{61}^2\bar{Y}^2A_{61} + K_{62}^2\bar{X}^2A_{62} - 2K_{61}K_{62}\bar{X}\bar{Y}A_{63} - 2K_{61}\bar{Y}^2A_{64} + \bar{Y}^2 + \frac{(L-1)}{n}KK_{61}^2S_{Y2}^2$$

(3.18)

Where $A_{61} = 1 + \theta\left[wC_1^2 + C_0^2 - 4w\rho C_0 C_1\right]$

$$A_{62} = \theta C_1^2$$

$$A_{63} = \theta\left[-wC_1^2 + \rho C_0 C_1\right]$$

$$A_{64} = 1 - \theta\left[\frac{w(w-1)}{2}C_1^2 - w\rho C_0 C_1\right]$$

And the optimum value of $K_{61}$ and $K_{62}$ is given as

$$K_{61} = \frac{A_{62}A_{64}}{A_{61}A_{62} - A_{63}^2 + \frac{(L-1)}{n}KA_{62}\frac{S_{Y2}^2}{\bar{Y}^2}}$$

$$K_{62} = \frac{\bar{Y}A_{63}A_{64}}{X\left[A_{61}A_{62} - A_{63}^2 + \frac{(L-1)}{n}KA_{62}\frac{S_{Y2}^2}{\bar{Y}^2}\right]}$$

**4. Empirical Study**

For numerical illustration, we have considered the data given in [18], The data are based on length (X) and timber volume (Y) for 176 forest strips. [12] and [18] reported the values of intraclass correlation coefficients $\rho_X$ and $\rho_Y$ approximately equal for the systematic sample of size 16 by enumerating all possible systematic samples after arranging the data in ascending order of strip length. The details of population parameters are: $N = 176$, $n = 16$, $\bar{Y} = 282.6136$, $\bar{X} = 6.9943$,

$S_Y^2 = 24114.6700$, $S_X^2 = 8.7600$, $\rho = 0.8710$, $S_{Y2}^2 = \frac{3}{4} S_Y^2 = 18086.0025$.

Table 6.1 shows the percentage relative efficiency (PRE) of $t^{**}$ (optimum) and $\bar{y}_{lr}^{**}$ with respect to $\bar{y}^{**}$ for the different choices of $K$ and $L$.

| Table 6. 1: PRE of estimators with respect to $\bar{y}^{**}$ | | | | | | | |
|---|---|---|---|---|---|---|---|
| K | L | PRE of $t_1$(optimum) with respect to $\bar{y}^{**}$ | PRE of $t_2$(optimum) with respect to $\bar{y}^{**}$ | PRE of $t_3$(optimum) with respect to $\bar{y}^{**}$ | PRE of $t_4$(optimum) with respect to $\bar{y}^{**}$ | PRE of $t_5$(optimum) with respect to $\bar{y}^{**}$ | PRE of $t_6$(optimum) with respect to $\bar{y}^{**}$ |
| 0.1 | 2.0 | 703.4864 | 407.4884 | 407.4884 | 419.8535 | 704.5781 | 840.4659 |
| | 2.5 | 692.3718 | 404.1824 | 404.1824 | 416.7079 | 687.6919 | 815.1533 |
| | 3.0 | 681.6592 | 400.9468 | 400.9468 | 413.6312 | 671.6886 | 791.4987 |
| | 3.5 | 671.3272 | 397.7794 | 397.7794 | 410.6211 | 656.5009 | 769.3449 |
| 0.2 | 2.0 | 681.6592 | 400.9468 | 400.9468 | 413.6312 | 671.6886 | 791.4987 |
| | 2.5 | 661.3558 | 394.6779 | 394.6779 | 407.6756 | 642.068 | 748.5538 |
| | 3.0 | 642.422 | 388.6647 | 388.6647 | 401.9702 | 615.2524 | 710.5873 |
| | 3.5 | 624.7238 | 382.8921 | 382.8921 | 396.5000 | 590.8619 | 674.8063 |
| 0.3 | 2.0 | 661.3558 | 394.6779 | 394.6779 | 407.6756 | 642.068 | 748.5538 |
| | 2.5 | 633.4262 | 385.7493 | 385.7493 | 399.2066 | 602.775 | 693.2089 |
| | 3.0 | 608.144 | 377.3458 | 377.3458 | 391.251 | 568.5821 | 644.7125 |
| | 3.5 | 585.15 | 369.4225 | 369.4225 | 383.7646 | 538.558 | 606.5479 |
| 0.4 | 2.0 | 642.422 | 388.6647 | 388.6647 | 401.8866 | 615.2524 | 710.5873 |
| | 2.5 | 608.144 | 377.3458 | 377.3458 | 391.251 | 568.5821 | 646.4959 |
| | 3.0 | 577.9409 | 366.881 | 366.881 | 381.3664 | 529.3474 | 594.4884 |
| | 3.5 | 551.1267 | 357.1773 | 357.1773 | 372.3468 | 495.9049 | 551.4543 |

## 5. Conclusion

In this paper, we have proposed general class of ratio-type, product-type and difference estimators for estimating the population mean in systematic sampling using auxiliary information in the presence of non-response. From the above empirical study we see the PRE of all estimators are decreasing with increasing non-response rate K as well as with increasing L. And here we see that in all proposed estimators, $t_6$ gives better result under non-response than other proposed estimators.